\documentstyle[prl,aps,multicol,epsf]{revtex}
\begin{document}
\draft
\title{Mimicking a turbulent signal: sequential multiaffine processes}
% \title{Mimicking a turbulent signal: sequential generation
% of multiaffine processes}

\author{L. Biferale}
\address{Dipartimento di Fisica, Universit\`a di Roma Tor Vergata,
Via della Ricerca Scientifica 1, 00133 Roma}
\address{and INFM Unit\`a di Tor Vergata, Italy}
\author{G. Boffetta}
\address{Dipartimento di Fisica Generale, Universit\`a di Torino,
         v. Pietro Giuria 1, 10125 Torino}
\address{and INFM Unit\`a di Torino, Italy}

\author{A. Celani}
\address{Dipartimento di Ingegneria Aeronautica e Spaziale,
	Politecnico di Torino,
	c. Duca degli Abruzzi 24, 10129 Torino}
\address{and INFM Unit\`a di Torino, Italy}

\author{A. Crisanti and A. Vulpiani}
\address{Dipartimento di Fisica, Universit\`a di Roma ``La Sapienza'',
         p.le Aldo Moro 2, 00185 Roma}
\address{and INFM Unit\`a di Roma I, Italy}

\date{\today}

\maketitle

\begin{abstract}
An efficient method for the construction of a multiaffine 
process, with prescribed  scaling exponents, is presented.
At variance with the previous proposals, this method is sequential
and therefore it is the natural candidate in numerical computations 
involving synthetic turbulence. 
The application to the realization of a realistic turbulent-like
signal is discussed in detail.
The method represents a first step towards the realization of a
realistic spatio-temporal turbulent field.
\end{abstract}

\pacs{05.40.+j,47.25.Cg,05.45.+b}
\begin{multicols}{2}

% ---------------------------------------------------------------

In recent years the relevance of multifractal measures and multiaffine
processes in many fields (mainly fully developed turbulence) has been 
well understood \cite{BPPV84,HJKPS86,PV87,PF85}. In different contexts,
for instance numerical simulations and comparison of theoretical models with 
experimental data, a rather natural problem is the construction of
artificial signals mimicking real phenomena (e.g. turbulence).
In particular it is important to have efficient numerical techniques
for the construction of a multiaffine field $\phi(x)$ whose structure 
functions scale as
\begin{equation}
\langle | \phi(x+r)-\phi(x) |^q \rangle \sim r^{\zeta_q}
\label{def:strfun}
\end{equation}
where $\langle \cdots \rangle$ indicates a spatial (or temporal) average,
$r$ varies in an appropriate scaling range and the exponents $\zeta_q$ are 
given. The most interesting case, and the most physically relevant, is when 
$\zeta_q$ is a nonlinear function of $q$, that is a strictly multiaffine 
field.

Let us first notice that the generation of a multiaffine function
is much more difficult that the generation of a multifractal measure,
which can be obtained with a simple multiplicative process generalizing
the two scales Cantor set.

Up to now, there exist well established methods for the construction 
of multiaffine fields \cite{VB91,EG92,BBCPVV93,JLSS94}, see \cite{JLSS94}
for a short review. 
All of these methods share the common characteristic of being not sequential:
the process is build as a whole in an interval (in space or time) of 
fixed length. To extend the interval one has to rebuild the process
from the beginning. This is an evident limitation if one is interested
in constructing a temporal signal mimicking, for example, those obtained by 
an anemometer measurement. Furthermore, non-sequential algorithms
require always a huge amount of stored data.

In this letter we introduce a simple and efficient sequential method for the
construction of a multiaffine function of time $u(t)$ with prescribed
statistical properties.
The guideline of our approach will be the reproduction of a 
turbulent-like temporal
signal. Though the basic idea on the construction of the multiaffine 
process comes from fully developed turbulence, nevertheless 
the method is general and can be applied to any signal.

A typical anemometer measurement gives a $1$-dimensional string
of data representing the one-point turbulent velocity $u(t)$ along the
direction of the mean flow $U$.
According to the Taylor 
hypothesis \cite{Taylor}, for small turbulence intensities
$u \ll U$, the time variations of $u$ can be assumed
to be due to the advection (with velocity $U$) 
of a frozen turbulent field past the measurement point, so that
\begin{eqnarray}
\delta u(\tau) & = & u(x,t+\tau) - u(x,t) = \nonumber \\
& = & u(x-U\tau,t) - u(x,t) = \delta u(\ell)
\label{taylor}
\end{eqnarray}
where $\ell = U \tau$. Therefore, once the spatial scaling  (\ref{def:strfun})
is given, we have:
\begin{equation}
S_{q}(\tau) = \langle | u(t+\tau)-u(t) |^q \rangle \sim \tau^{\zeta_q} \; .
\label{def:strfun2}
\end{equation}
The frozen field is the result of the superposition of turbulent 
patterns (eddies) of many different sizes $\ell$, whose 
contribution to the time variation of the velocity 
decays with a typical correlation time $\tau_{sweep}\sim \ell/U$.
For the sake of simplicity, in the following,  we shall introduce a set
of reference scales $\ell_n = 2^{-n}$ at which scaling properties will be 
tested. 
With this picture in mind, we represent the signal $u(t)$ by a 
superposition of functions with different characteristic times, 
representing eddies of various sizes
\begin{equation}
u(t)=\sum_{n=1}^N v_n(t) \; .
\label{eq:decomp}
\end{equation}
The functions $v_n(t)$ are defined via a multiplicative process
\begin{equation}
v_n(t)=g_n(t)x_1(t)x_2(t)\ldots x_n(t) \; ,
\label{def:mult}
\end{equation}
where the $g_n(t)$ are independent stationary random processes, whose 
correlation times are the sweeping timescales 
$\tau_n=\ell_n/U=2^{-n}$ (assuming $U=1$) and 
$\langle g_n^2 \rangle = \ell_n^{\,2h}$
where $h$ is the scaling exponent. For fully developed turbulence $h=1/3$. 
Scaling will show up for all time delay larger than the UV 
cutoff $\tau_N$ and smaller than the IR cutoff $\tau_1$.
The $x_j(t)$ are independent, positive defined, identical distributed
random processes whose time correlation decays with characteristic
time $\tau_j$. The probability distribution of $x_j$ determines the
intermittency of the process.

The origin of (\ref{def:mult}) is fairly clear in the context of fully
developed turbulence. Indeed according to the Refined Similarity Hypothesis
of Kolmogorov \cite{K62,F95}, we can identify $v_n$ with the velocity 
difference at scale $\ell_n$ and $x_j$ with 
$(\varepsilon_j/\varepsilon_{j-1})^{1/3}$, where
$\varepsilon_j$ is the energy dissipation at scale $\ell_j$.

It is easy to show, with a simple argument, that the process 
constructed according to (\ref{eq:decomp},\ref{def:mult}) is multiaffine.
Because of the fast decrease of the correlation times $\tau_j=2^{-j}$,
the characteristic time of $v_n(t)$ is of the order of the shortest one, 
i.e., $\tau_n=2^{-n}$.
Therefore, the leading contribution to the structure function
$S_q(\tau)$ with $\tau \sim \tau_n$ will stem from the $n$-th term in 
(\ref{eq:decomp}). This can be understood nothing that in the sum 
$
u(t+\tau)-u(t) = \sum_{k=1}^N [v_k(t+\tau)-v_k(t)]
$ the terms with $k \le n$ are negligible because $v_k(t+\tau) \simeq v_k(t)$
and the terms with $k \ge n$ are subleading.
Thus one has:
\begin{equation}
S_q(\tau_n) 
\sim \langle |v_n|^q \rangle \sim 
\langle |g_n|^q \rangle \langle x^q \rangle^n
\sim \tau_n^{hq-\log_2\langle x^{q} \rangle } 
\end{equation}
and therefore for the scaling exponents (\ref{def:strfun2})
\begin{equation}
\zeta_q=hq-\log_2\langle x^{q} \rangle \; .
\label{eq:zq}
\end{equation}
The limit of an affine function can be obtained when 
all the $x_j$ are equal to $1$.

The above results can be proved in a more rigorous way considering the second
order structure function $S_2(\tau)$.
Using the definitions (\ref{eq:decomp},\ref{def:mult}) and stochastic 
independence one obtains:
\begin{equation}
S_2(\tau)= 2 \displaystyle{\sum_{n=1}^N} [
 \langle v_n(t)^2 \rangle - \langle v_n(t)v_n(t+\tau) \rangle ].
\label{eq:s2}
\end{equation}
Let us now introduce the normalized correlation functions for 
$g_n(t)$ and $x_j(t)$
\begin{equation}
C\left({s \over \tau_n}\right)=\displaystyle{
{\langle g_n(t+s)g_n(t) \rangle \over \langle g_n^2 \rangle }
} 
\end{equation}
\begin{equation}
F\left({s \over \tau_j}\right)=\displaystyle{
{\langle x_j(t+s)x_j(t) \rangle \over \langle x_j^2 \rangle }
}
\end{equation}
Plugging into (\ref{eq:s2}) the definition (\ref{def:mult}) one obtains
\begin{equation}
S_2(\tau)=2\sum_{n=1}^N \langle g_n^2 \rangle
\langle x^2 \rangle^n \left(
1-C({\tau \over \tau_n}) F({\tau \over \tau_1})
\cdots F({\tau \over \tau_n})  \right) \; .
\label{eq:s2bis}
\end{equation}
By shifting the summation index in the above expression,
$n \rightarrow n-1$, one obtains for $\tau \ll 1$,
\begin{equation}
S_2(2\tau)\sim 2^{2h} \langle x^2 \rangle^{-1} S_2(\tau)
\label{eq:scaling}
\end{equation}
which leads to the scaling behavior
\begin{equation}
\begin{array}{lcr}
S_2(\tau) \sim \tau^{\zeta_2} & \hspace{20pt} \mbox{with} \hspace{10pt} &
\zeta_2 = \displaystyle{2h}-\log_{2} \langle x^2 \rangle \; .
\end{array}
\end{equation}
A similar computation can be performed for the higher order structure 
functions. The generic 
$S_q(\tau)$ can be expressed as a linear combination of terms 
scaling as $\tau^{\zeta_{m_1}}\cdots\tau^{\zeta_{m_k}}$ with 
$m_1+\ldots+m_k=q$. From the convexity of  $\zeta_q$ \cite{Feller}
it follows that the leading contribution to $S_q(\tau)$ for small $\tau$ 
is given by $S_q(\tau) \sim \tau^{\zeta_q}$, with the exponents
$\zeta_q$ as defined in (\ref{eq:zq}).

The key point in the above arguments is that the dominant contribution
to the structure function $S_q(\tau)$ comes from octaves $n$
such that $\tau_n \sim \tau$, that is locality.

The constraints for locality can be captured with a simple argument.
Indeed for $\tau_n \ll \tau$ we have that 
$\langle |v_n(t+\tau)-v_n(t)|^q \rangle \sim \langle |v_n|^q \rangle 
\sim 2^{-n\zeta_q}$, therefore UV convergence requires $\zeta_q >0 $.
Similarly, when  $\tau_n \gg \tau$ we have that:
$ \langle |v_n(t+\tau)-v_n(t)|^q \rangle \sim 
(\tau/\tau_n)^{q/2} \langle|v_n|^q\rangle \sim 2^{-n(\zeta_q-q/2)}$, 
for   stochastic processes
such that $C(x) = 1 -O(x)$ and  $F(x) = 1-O(x)$. Therefore,
convergence in the latter case 
 requires  $\zeta_q < q/2$.
We observe that the last condition is different from the usual locality
condition $\zeta_q < q$ \cite{Rose}
which holds for differentiable processes where 
$C(x) = 1-O(x^2)$ and $F(x) = 1-O(x^2)$. 

Regular behavior for very short time delays
$\delta u(\tau) \sim \tau$, 
physically related to the presence of dissipation, 
can be simply achieved in our model by smoothing 
$g_n(t)$ and $x_n(t)$ over a time interval smaller then the UV
cutoff $ \tau_N$.

The numerical implementation of the method proposed above is very simple.
The stochastic process $x_j(t)$ can be easily generated via
the nonlinear Langevin differential equations:
\begin{equation}
dx_j=-{1 \over \tau_j}{dV \over dx_j}dt+\sqrt{{2 \over \tau_j}} \, dW_j \; 
\label{eq:sde}
\end{equation}
where $V(x)=\infty$ for $x<a$ ($a$ positive constant) and 
$V(x) \rightarrow \infty$ for $x \rightarrow \infty$.
It is clear that the $x_j$ so obtained have the same 
probability density function independently of $\tau_j$.

Similarly for the $g_n$ one can use the evolution law
\begin{equation}
dg_n=-{1 \over \tau_n}{dY \over dg_n}dt+
\sigma_n\sqrt{{2 \over \tau_n}} \, dW_n \; ,
\label{eq:sdeg}
\end{equation}
where $Y(g) \rightarrow \infty$ as $|g| \rightarrow 
\infty$ and $\sigma_n=\ell_n^h$.

Numerical tests have been performed 
adopting for the stochastic differential equations
(\ref{eq:sde},\ref{eq:sdeg}) the following potentials:
\begin{equation}
V(x) = -2 \ln x \qquad \mbox{for} \ (1-b)^{1/3}<x<(1+b)^{1/3} 
\label{eq:vx}
\end{equation}
and $V(x)=\infty$ otherwise, where  $0<b<1$ , while
\begin{equation}
Y(g) = {1 \over 2}g^2 \; .
\end{equation}
For $h=1/3$, this choice insures that $\zeta_3=1$ according to the scaling 
prescribed by Kolmogorov's law. The parameter $b$ tunes the intermittency
of the signal: when $b \downarrow 0$ we recover an affine process.  

In figure 1 we show the the quantity 
$v_N^2(t)$
which can be considered as the energy density dissipation of the
turbulent signal. As one can see high intermittency is detected. 

The theoretical and numerical scaling laws are compared in figure 2.
The computed scaling exponents are in perfect agreement 
with those given by equation (\ref{eq:zq}).
Figure 3 shows the probability density function of the velocity differences
$\delta u(\tau)=u(t+\tau)-u(t)$ for different $\tau$.
At large $\tau \sim 1$ the pdf is nearly Gaussian, whereas
at small delays the pdf is increasingly peaked around zero
with high tails corresponding to large fluctuations with
respect to their rms value.
 If one wants the process $u(t)$ to have a 
nonzero skewness, as in turbulence, $Y(g)$ must be chosen as an
asymmetric function, see \cite{JLSS94} for a suitable choice 
according to experimental data.

In this letter we have introduced an  efficient sequential
algorithm for the generation of multiaffine processes.
This method, at variance with previous proposals,
is not based on hierarchical construction,
and can be applied to any multiaffine signals
with specified scaling laws. Furthermore, no huge 
amount of memory is required for the numerical implementation.

A possible, relevant, application of such a signal would
be to use it for describing the temporal part
of a synthetic turbulent velocity field. The spatial part
can be implemented by using any hierarchical constructions
previously proposed \cite{VB91}-\cite{JLSS94}.
Nevertheless, this way to glue together spatial and temporal
multiaffine fluctuations would not be realistic, due to 
the absence of a real sweeping of small scales by large
scales. This is connected to the fact that in our temporal
signal, the Taylor hypothesis is introduced by hands
without any real direct dynamical (stochastic) 
coupling between large and small
scales.

These difficulties in reproducing an Eulerian spatio-temporal
field are absent if one considers the velocity statistics 
in quasi-lagrangian coordinates \cite{pro}. 
In this framework a pure temporal signal would correspond 
to the velocity field felt in the moving reference frame attached
to a fluid particle.
The sweeping effect is thus removed and the characteristic time
scales are the dynamical eddy turnover times.
Work in this direction is in progress.

We thank D. Pierotti for useful discussions in the early
stage of the work. 
This work has been partially supported by the INFM 
(Progetto di Ricerca Avanzata TURBO).

% ------------------------------------------------------------------

% ---------------------------------------------------------------
\narrowtext

\begin{figure}[hbt]
\epsfxsize=210pt\epsfysize=186.68pt\epsfbox{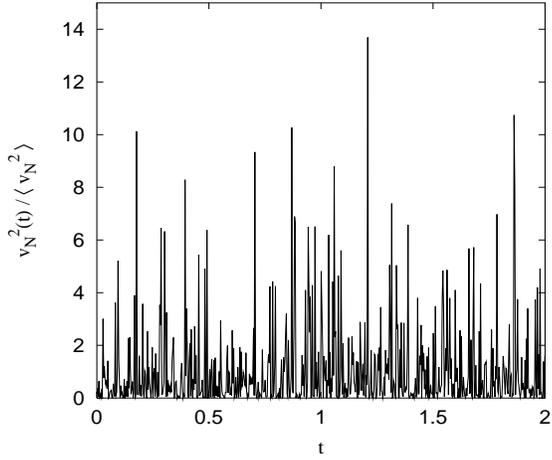}
\caption{Time series $v_N^2(t)$ normalized to the average 
for the model with $N=15$ octaves and $b=0.9$.}
\label{figure1}
\end{figure} 

\vbox{
\begin{figure}[hbt]
\epsfxsize=210pt\epsfysize=186.68pt\epsfbox{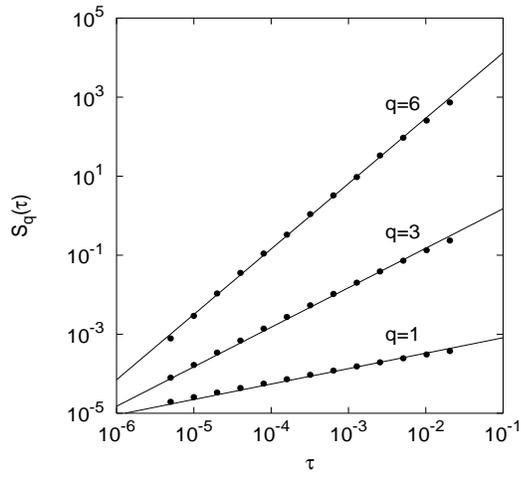}
\caption{Numerical (dots) and theoretical (line) structure functions
$S_{q}(\tau)$ for the  model with $N=20$ octaves and $b=0.9$.
The exponents are $\zeta_1=0.39$,
$\zeta_3=1$,$\zeta_6=1.65$. The structure functions are shifted 
by a multiplicative factor for plotting purposes.}
\label{figure2}
\end{figure} 
}

\begin{figure}[hbt]
\epsfxsize=210pt\epsfysize=186.68pt\epsfbox{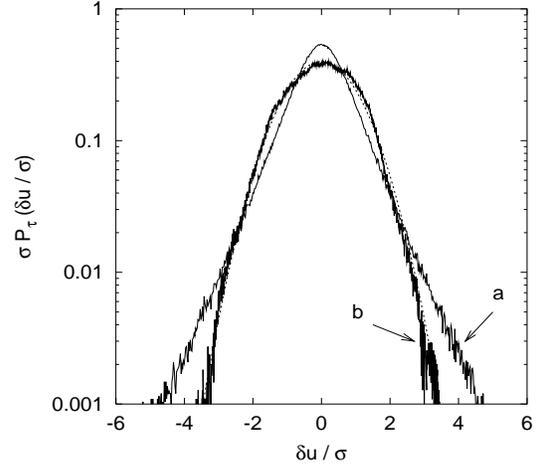}
\caption{Probability density functions for the normalized velocity 
differences $\delta u (\tau)/\sigma$, where 
$\sigma =\langle \delta u^2 \rangle^{1/2}$,  for differents $\tau$.
For large $\tau=10$ (b) the pdf is nearly Gaussian
(dashed curve). For very small $\tau=0.001$ (a)
large tails are evident.
The parameters are $N=15$ octaves and $b=0.9$.
}
\label{figure3}
\end{figure} 

\end{multicols}
\end{document}